\input epsf.tex
\input xpt.def
\input xivpt.def
\input xipt.def
\hsize=5in \hoffset=1.7cm
\vsize=8in \voffset=2.0cm
\headline={\it Kopacki {\em et al.} -- Variable stars in NGC 7044 \hfil\rm\folio}
\footline={\hfil}
\parskip=2pt plus2pt minus2pt
\let\em=\it
\let\normal=\xipt
\let\small=\xpt
\let\large=\xivpt
\def\,{\hskip2pt}
\def\begintfig#1{%
 \topinsert
 \hbox to\hsize{\hfill\epsfbox{#1}\hfill}
 \small\rm
 \noindent}
\def\beginmfig#1{%
 \midinsert
 \hbox to\hsize{\hfill\epsfbox{#1}\hfill}
 \small\rm
 \noindent}
\def\endfig{%
 \endinsert
 \normal\rm
 }

\hbox{}\vskip2cm

\centerline{\large\bf Variable Stars in the Open Cluster NGC\,7044}

\vskip0.3cm
\centerline{by}

\vskip0.3cm
\centerline{G.\ K o p a c k i$^{\ 1}$,  D.\ D r o b e k$^{\ 1}$,}
\centerline{Z.\ K o \l{} a c z k o w s k i$^{\ 1,2}$ and G.\ P o \l{} u b e k$^{\ 1}$}

\vskip0.3cm
\centerline{$^{1}$Instytut Astronomiczny Uniwersytetu Wroc\l{}awskiego,}
\centerline{Kopernika 11, 51-622 Wroc{\l}aw, Poland}

\centerline{$^{2}$Departamento de F\'\i{}sica, Universidad de Concepci\'on,}
\centerline{Casilla 160-C, Concepci\'on, Chile}

\centerline{E-mail: (kopacki,drobek)@astro.uni.wroc.pl}

\vskip0.5cm
\centerline{\it Received...}

\vskip1cm
\centerline{ABSTRACT}

\vskip0.5cm
\small

\noindent
We present results of a search for variable stars in the intermediate-age
open cluster NGC\,7044. We found 23 variable stars in the observed field.
One star turned out to be of the $\delta$ Sct type with two pulsational
modes excited. From the position in the color-magnitude diagram we conclude
that this star is a member of the cluster. Moreover, we found 13 eclipsing
systems, of which five are W UMa stars, one is a $\beta$ Lyr variable,
six are $\beta$ Per binaries showing detached configuration, and the
last one is another probable $\beta$ Per system.
Using the period-luminosity-color relation for W UMa stars we established 
the membership of the contact binaries, finding four of them 
to be very probable cluster members. We estimated from these four stars an apparent
distance modulus $(m-M)_V$ of NGC\,7044 to be $14.2\pm0.4$ mag, which is smaller
than previous determinations of this parameter.
We were able to derive orbital period for only four $\beta$ Per systems. For 
the remaining ones we observed only two or three eclipses. Finally, nine  
stars we found to show irregular light changes. Most of them are red stars not
belonging to the cluster.
For the cluster core we determined a reddening map, which allowed
us to construct a dereddened color-magnitude diagram of NGC\,7044 with
a narrow main-sequence. By
fitting a theoretical isochrone to this diagram we derived
$E(V-I_{\rm C})={}$0.92 mag, $(m-M)_V={}$14.45 mag and
$\log(\tau/\rm{yr})={}$9.2.

\vskip0.2cm


\normal

\def\figfld{1}
\def\figdsctspectrum{2}
\def\figdsctlc{3}
\def\figwumalc{4}
\def\figwumaplr{5}
\def\figdeblc{6}
\def\figdebvxii{7}
\def\figdebvxiii{8}
\def\figdebvxiv{9}
\def\figrvlc{10}
\def\figcmd{11}
\def\figcmdredmap{12}
\def\tabvarpar{1}

\def\secIntro{1}
\def\secObsRed{2}
\def\secVar{3}
\def\secVarDSct{3.1}
\def\secVarWUMa{3.2}
\def\secVarBPer{3.3}
\def\secVarIrr{3.4}
\def\secCMD{4}
\def\secRMFit{5}

\vskip0.5cm
\centerline{\bf \secIntro. Introduction}
\vskip0.5cm

NGC\,7044 ($\alpha=21^{\rm h}13^{\rm m}09^{\rm s}$, 
$\delta=42^\circ29^\prime42^{\prime\prime}$) is an intermediate-age open cluster located 
in a dense region of the Milky Way in Cygnus. Because of its
faintness it remained unstudied for a very long time. The first modern 
study of NGC\,7044 was published by Ka\l{}u\.zny (1989). 
From the $BV$ CCD photometry, Ka\l{}u\.zny (1989) determined the reddening,
$E(B-V)=0.74$ mag, an approximate distance of about 4 kpc and the age of
about 1.5 Gyr. The next CCD observations of NGC\,7044
were obtained by Aparicio {\em et al}.\ (1993). They derived different
cluster parameters: $E(B-V)=0.57$ mag, distance 3 kpc and age 2.5 Gyr.
Moreover, from the color-color diagram they estimated cluster metallicity [Fe/H]${}=0$. Finally,
Sagar and Griffiths (1998a) presented $BVI$ observations of NGC\,7044
and obtained $E(B-V)=0.7$ mag, $E(V-I)=0.88$ mag, the same distance
as Aparicio {\em et al}.\ (1993) and an age of 1.6 Gyr by isochrone
fitting to the cluster color-magnitude diagram.

Ka{\l}u\.zny (1989) and Sagar and Griffiths (1998a) noticed relatively wide
main sequence of NGC~7044 and attributed its origin to
variable reddening or large number of binaries in the cluster field.
The mass function of NGC\,7044 and four other open clusters
was determined and studied by Sagar and Griffiths (1998b).
A comparison of all photometric observations mentioned above was done by
Sagar and Griffiths (1998a). They concluded that the data of
Aparicio {\em et al}.\ (1993) differ systematically both in color and 
magnitude from the observations of Ka\l{}u\.zny (1989) and
Sagar and Griffiths (1998a) which agree with each other.

Until now the cluster was not a subject of a search for variable stars. 
In this paper we present our time-series observations of NGC\,7044 and
announce discovery of 23 variable stars in the cluster field. 

\vskip0.7cm
\centerline{\bf \secObsRed. Observations and Reductions}
\vskip0.5cm

The observations presented here were obtained at the
Bia\l{}k\'ow Observatory of the University of Wroc\l{}aw using
a 60-cm Cassegrain telescope equipped with Andor DW432-BV
back-illuminated CCD camera. One 12.8${}\times{}$11.7 
arcmin$^{2}$ field centered on 
NGC\,7044 was observed through $V$ and $I_{\rm C}$ filters
of the Johnson-Kron-Cousins $UBV(RI)_{\rm C}$ photometric
system.

The observations were carried out on 26 nights between 
2005 August 4 and September 25. Because we did not have 
an autoguider, the exposure times were rather short, amounting 
to only 100 s. In total, we collected 1924 and 703 CCD frames of
NGC\,7044 in the $V$ and $I_{\rm C}$ passbands, respectively.
On most nights the weather was very good. The seeing changed 
over a rather wide range, with a typical value of 2.5 arcsec.

\begintfig{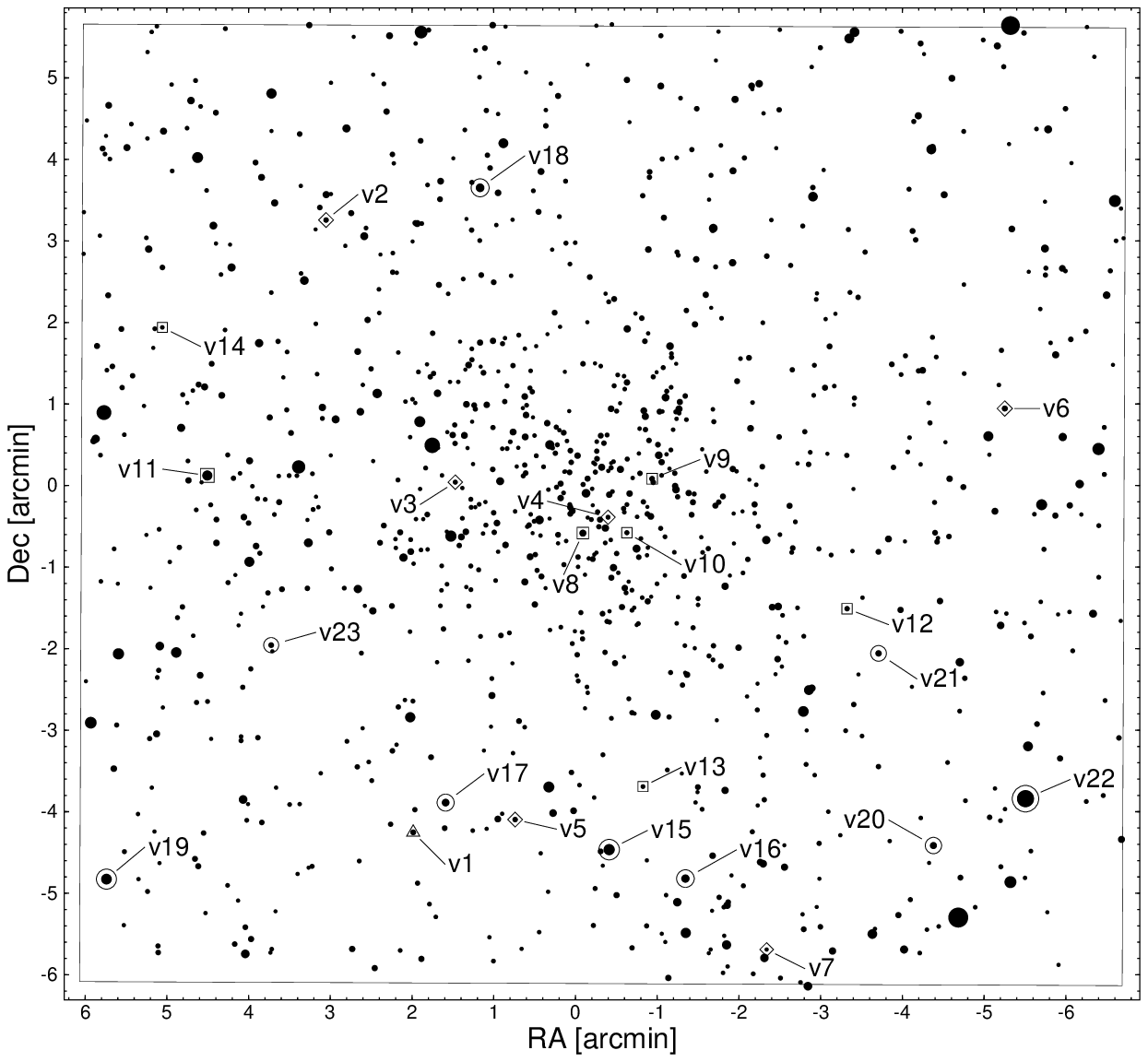}%
 Fig.\ \figfld. Schematic view of the observed field of
 NGC\,7044. For clarity, only stars brighter than 18 mag in $V$ are shown
 (filled circles with diameter proportional to brightness).
 Variable stars are additionally indicated with open symbols,
 one $\delta$ Sct star, with a triangle, W UMa and $\beta$ Lyr 
 binary systems, with diamonds, $\beta$ Per systems, with
 squares, and irregular variable stars with circles.
 Equatorial coordinates of the $(0,0)$ point are 
 $\alpha=21^{\rm h}13^{\rm m}09^{\rm s}$, 
 $\delta=42^\circ29^\prime42^{\prime\prime}$.
\endfig

The pre-processing of the frames was performed in the usual 
way and consisted of subtracting bias and dark frames and
applying the flat-field correction. The $I_{\rm C}$-filter
frames were also corrected for the fringing pattern.
Instrumental magnitudes for all stars in the field
were computed using the DAOPHOT profile-fitting software
(Stetson 1987). All images were reduced in the same way 
as described by Jerzykiewicz {\em et al}.\ (1996).
We identified 4800 stars in the $V$-band reference frame 
of the observed field, but only about 2800 of them had
photometry of reasonable quality. A finding chart for the 
field we monitored is shown in Fig.\ \figfld. For clarity,
we show only stars brighter than 18 mag in $V$.

The differential photometry was derived on frame-to-frame basis
rather than on the usual star-to-star basis. That is,
instrumental photometry for each frame was shifted to the same
magnitude scale (defined by means of a reference frame) 
by an average offset in brightness between a
given frame and the reference frame. The average offset was determined
from a large number of bright
unsaturated stars; $\sigma$-clipping was applied in order to 
reject deviating stars. In this way slightly different set of comparison
stars was used for each frame, but the magnitude offsets
are very accurate because they were computed as mean values.

Our average instrumental magnitudes and colors were
transformed to the standard system using the
photometric data of Sagar and Griffiths (1998a).
From 760 bright stars in common with Sagar and Griffiths 
(1998a) we obtained the following transformation 
equations:

\vskip0.7\baselineskip
 \hbox to\hsize{\hfill\vbox{\tabskip=1pt 
  \halign{%
   \hfil#\tabskip=5.5pt&%
   #\hfil\tabskip=1.5cm&%
   #\hfil\tabskip=1pt\cr
   $V-v$&         = $-0.027\,(v-i)-0.751$,& $\sigma={}$0.031,\cr
%  $I_{\rm C}-i$& = $-0.022\,(v-i)-1.260$,& $\sigma={}$0.037,\cr
   $V-I_{\rm C}$& = $+0.995\,(v-i)+0.509$,& $\sigma={}$0.039,\cr
  }}\hfill}
\vskip0.4\baselineskip

\noindent
where uppercase letters denote standard magnitudes
and lowercase letters, the instrumental magnitudes;
$\sigma$ is the standard deviation of the fit.

Individual instrumental $V$- and $I_{\rm C}$-filter  magnitudes of the
periodic variable stars were transformed to the standard system
in the following way.
For each variable and both filters, phased light-curves were
decomposed into Fourier series with the number of harmonics
fixed at a value that gave a satisfactory fit. The quality of
the fit was judged by eye.
%chosen by eye to obtain the best fit.
Next, instrumental magnitudes were
transformed to the standard ones using the above-given equations
and the color indices defined
as a difference of these two smooth curves
at the appropriate phase of variation. 

In order to search for periodic variable stars in 
the observed field, for each star we computed Fourier
amplitude spectrum in the frequency range from 0 to 40 d$^{-1}$.
We looked at the light curves of all stars with significant signal-to-noise
ratio of the hightest peak in the spectrum. Moreover,
the light curves of all bright stars were checked manually for
the presence of eclipses. This analysis resulted in 
finding 23 variable stars not known previously in the field
of NGC\,7044.

\vskip0.7cm
\centerline{\bf \secVar. Variable stars}
\vskip0.5cm

\def\mp{\omit\hfil--\hfil}
%\vskip1.0\baselineskip plus3pt minus1pt
\topinsert
\vbox{\noindent\hbox to\hsize{\hss T a b l e \tabvarpar\hss}\par
\vskip0.3\baselineskip\small
\noindent\hfil Photometric data for variable stars in \hbox{NGC\,7044}\par
\vskip\baselineskip
\normalbaselineskip=\baselineskip
\setbox\strutbox=\hbox{\vrule height0.7\normalbaselineskip%
 depth0.3\normalbaselineskip width0pt}
%\normalbaselines
 \hbox to\hsize{\hfill\vbox{\tabskip=1pt 
  \halign{%
   \hfil#\hfil\tabskip=5.5pt&%
   \hfil#\hfil&%
   \hfil#\hfil&%
   \hfil#\hfil&%
   \hfil#\hfil&%
   \hfil#\hfil&%
   #\hfil&%
   #\hfil&
   #\hfil%
   \tabskip=1pt\cr
%   \noalign{\hrule\vskip1.7pt}
   \noalign{\hrule\vskip3pt}
   Var& Type& 
   $\alpha_{2000}$& $\delta_{2000}$& 
   $V$& $V$-$I_{\rm C}$& \omit\hfil$\Delta V$\hfil& \omit\hfil$\Delta I_{\rm C}$\hfil&
   \omit\hfil$P$\hfil\cr
   \noalign{\vskip1pt}
     && [$^{\rm h}$ $^{\rm m}$ $^{\rm s}$]& [$^\circ$ $^\prime$ $^{\prime\prime}$]&
     [mag]& [mag]& \omit\hfil[mag]\hfil& \omit\hfil[mag]\hfil& \omit\hfil[d]\hfil\cr
   \noalign{\vskip3pt\hrule\vskip3pt}
%  no   Type         ra         dec             V   (V-I)  DV  DI        P     
   v1& $\delta$ Sct& 21 13 19.74& 42 25 26.8& 16.838& 1.423& 0.029& 0.022& 0.088957\cr
                                                      &&&&&& 0.017&   \mp& 0.091464\cr   
%                                                     &&&&&& 0.018& 0.007& 0.091461\cr
%                                                     &&&&&& 0.010&   \mp& 0.069467\cr
  \noalign{\vskip2pt}
   v2& W UMa& 21 13 25.56& 42 32 57.4& 17.158& 1.413& 0.117& 0.105& 0.33388\cr
   v3& W UMa& 21 13 16.97& 42 29 44.5& 17.642& 1.414& 0.421& 0.397& 0.46057\cr
   v4& W UMa& 21 13 06.83& 42 29 18.7& 17.745& 1.558& 0.559& 0.535& 0.50363\cr
   v5& W UMa& 21 13 12.99& 42 25 36.2& 17.097& 1.447& 0.470& 0.456& 0.61501\cr
   v6& W UMa& 21 12 40.49& 42 30 38.7& 16.616& 1.411& 0.269& 0.263& 0.65472\cr
  \noalign{\vskip2pt}
   v7& $\beta$ Lyr& 21 12 56.32& 42 24 00.6& 18.504& 1.281& 0.393& 0.349& 0.87539\cr
  \noalign{\vskip2pt}
   v8& $\beta$ Per& 21 13 08.50& 42 29 07.0& 15.790& 2.034& 0.12& 0.08& 1.83882\cr
   v9& $\beta$ Per& 21 13 03.90& 42 29 47.0& 16.944& 1.450& 0.39& 0.37& 2.5269\cr
  v10& $\beta$ Per& 21 13 05.58& 42 29 07.3& 17.505& 1.622& 0.17& 0.18& 2.9056\cr
  v11& $\beta$ Per& 21 13 33.43& 42 29 49.4& 13.943& 1.205& 0.09& 0.08& 4.1528\cr
  v12& $\beta$ Per& 21 12 50.97& 42 28 11.4& 17.174& 1.443& 0.52& 0.27& \mp\cr
  v13& $\beta$ Per& 21 13 04.52& 42 26 00.5& 17.940& 1.677& 0.26& 0.11& \mp\cr
  \noalign{\vskip2pt}
  v14& $\beta$ Per& 21 13 36.43& 42 31 38.4& 18.158& 1.949& 0.43& 0.27& \mp\cr
  \noalign{\vskip2pt}
% v15& Irr& 21 13 17.26& 42 29 04.8& 13.120& 2.740& 0.03& 0.02& \mp\cr
  v15& Irr& 21 13 06.76& 42 25 14.1& 13.406& 3.248& 0.09& 0.06& \mp\cr
  v16& Irr& 21 13 01.70& 42 24 52.8& 14.932& 2.929& 0.08& 0.05& \mp\cr
  v17& Irr& 21 13 17.60& 42 25 48.7& 15.378& 1.055& 0.06& 0.04& \mp\cr
  v18& Irr& 21 13 15.33& 42 33 21.1& 14.721& 3.260& 0.10& 0.05& \mp\cr
  v19& Irr& 21 13 40.09& 42 24 52.4& 13.587& 3.785& 0.20& 0.07& \mp\cr
% v21& Irr& 21 13 03.66& 42 26 53.3& 13.679& 2.922& 0.03& 0.02& \mp\cr
% v22& Irr& 21 13 19.33& 42 30 29.0& 13.166& 2.663& 0.03& 0.02& \mp\cr
% v23& Irr& 21 13 16.27& 42 29 08.0& 15.601& 1.749& 0.03& 0.03& \mp\cr
  v20& Irr& 21 12 45.26& 42 25 17.1& 15.759& 3.747& 0.25& 0.09& \mp\cr
  v21& Irr& 21 12 48.88& 42 27 38.5& 16.200& 5.641& 0.19& 0.15& \mp\cr
  v22& Irr& 21 12 39.15& 42 25 51.6& 11.326& 3.009& 0.20& 0.09& \mp\cr
  v23& Irr& 21 13 29.18& 42 27 44.6& 16.840& 5.264& 0.70& 0.41& \mp\cr
  \noalign{\vskip3pt\hrule}
  }}\hfill}}
\endinsert
%\vskip0.5\baselineskip plus3pt minus1pt

Among the 23 variable stars we detected there is
one $\delta$ Sct star, 13 eclipsing systems and nine irregular variables.
On the basis of the period and  the light-curve shape we distinguished
among eclipsing systems five W UMa stars, one $\beta$ Lyr system,
six $\beta$ Per binaries showing indication of detached configuration and one possible
$\beta$ Per system.
We denote these stars v1 through v23. Their photometric parameters
are given in Table \tabvarpar.
For each variable we provide the adopted designation, type of variability, 
equatorial coordinates, $(\alpha,\delta)$, mean brightness in $V$, 
mean color index $(V-I_{\rm C})$, the ranges of variability, $\Delta V$ and
$\Delta I_{\rm C}$, and where possible, period(s), $P$.
Periods are given with an accuracy resulting from a non-linear least-squares
fit of truncated Fourier series to the observations.

\vskip0.6\baselineskip plus1pt minus1pt
{\em \secVarDSct. The $\delta$ Sct star}
\vskip0.4\baselineskip plus1pt minus1pt

Star v1 shows sinusoidal brightness variations
in both filters, with the amplitude in $V$ greater than 
in $I_{\rm C}$. Fourier analysis of the $V$-filter observations  
revealed frequency $f_1=11.241$ d$^{-1}$ (see Fig.
\figdsctspectrum{}a). After prewhitening the data with $f_1$, another
peak, corresponding to frequency $f_2=10.934$ d$^{-1}$ appeared in
the amplitude spectrum (see Fig. \figdsctspectrum{}b). It should be 
noted that due to the daily aliasing, the secondary frequency 
may be incorrect by 1 d$^{-1}$. The strongest alias of $f_2$ frequency, 
$f_2+1$ d$^{-1}$, has almost the same amplitude. Fitting the sum
of sine terms with $f_1$ and $f_2$ or its alias to the original data
gives almost the same standard deviation of $0.027$ mag.
It is therefore difficult to conclude which secondary frequency is the 
true one. Further observations of this star are needed to resolve this ambiguity.
We may, however, safely classify v1 as a $\delta$ Sct star showing 
at least two frequencies.

\begintfig{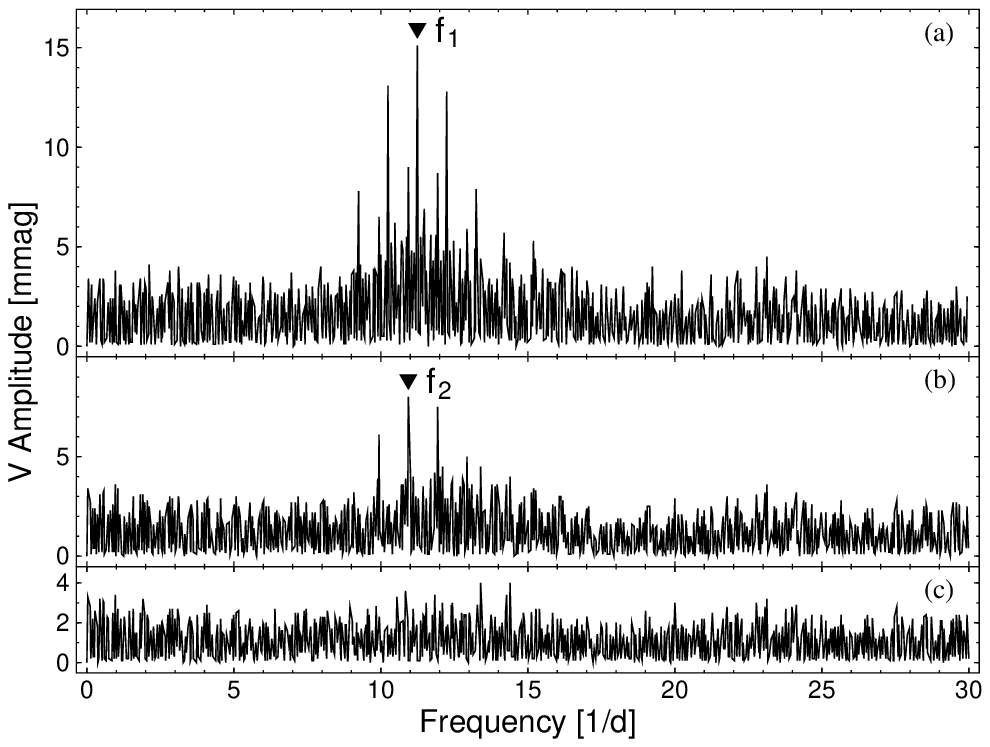}%
 Fig.\ \figdsctspectrum. Fourier amplitude spectra of the $\delta$ Sct type star v1: 
 (a) for original $V$-filter observations,
 (b) after prewhitening with frequency $f_1=11.241$ d$^{-1}$, and
 (c) after removing terms with frequencies $f_1$ and $f_2=10.934$ d$^{-1}$.
%and (d) after prewhitening with frequencies $f_1$, $f_2$, and $f_3=14.395$ d$^{-1}$.
 The ordinate scale is the same in {\em all panels}.
\endfig

\begintfig{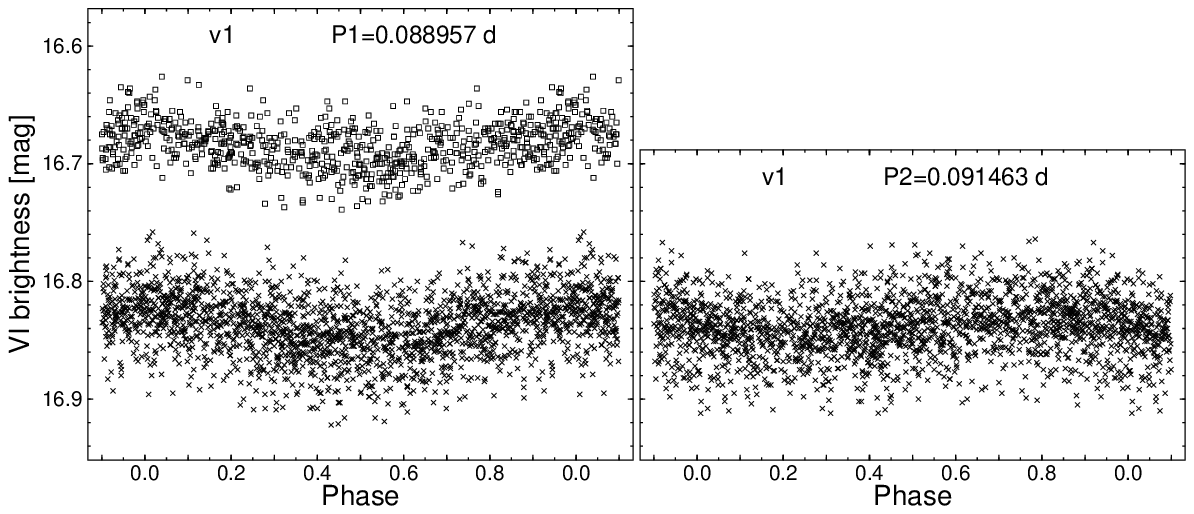}%
 Fig.\ \figdsctlc. $V$ (crosses) and $I_{\rm C}$ (squares) light-curves 
 of the $\delta$ Sct star v1.
 The {\em left panel} shows the residuals obtained after removing the
 $f_2$ component, phased with the period $P_1$.
 The {\em right panel} shows the data prewhitened with 
 $f_1$ and phased with the period $P_2$.
 Initial epoch was chosen arbitrarily, but it
 is the same for each light curve.
 The ordinate scale is the same in {\em both panels}.
\endfig

The amplitudes of both sinusoidal components detected in v1 are given
in Table \tabvarpar. We were able to find only the primary frequency 
in the $I_{\rm C}$-filter observations of this star. The phase
diagrams of the two pulsational components of v1 are shown 
in Fig.\ \figdsctlc.

Judging from the position of v1 % in the upper part of the main sequence
in the color-magnitude diagram (see Fig.\ \figcmd) we may assume
that the star is a cluster member. Using this fact we can determine its absolute magnitude.
We adopted the values of distance modulus $(m-M)_V$ and reddening $E(B-V)$ for NGC\,7044
from Ka\l{}u\.zny (1989) and Sagar and Griffiths (1998a).
In both cases the resulting absolute brightness $M_V$ of v1 (1.5 mag and 2.2 mag,
respectively) places the star in the classical instability strip. This
further supports our
classification of v1 as a $\delta$ Sct star.

\begintfig{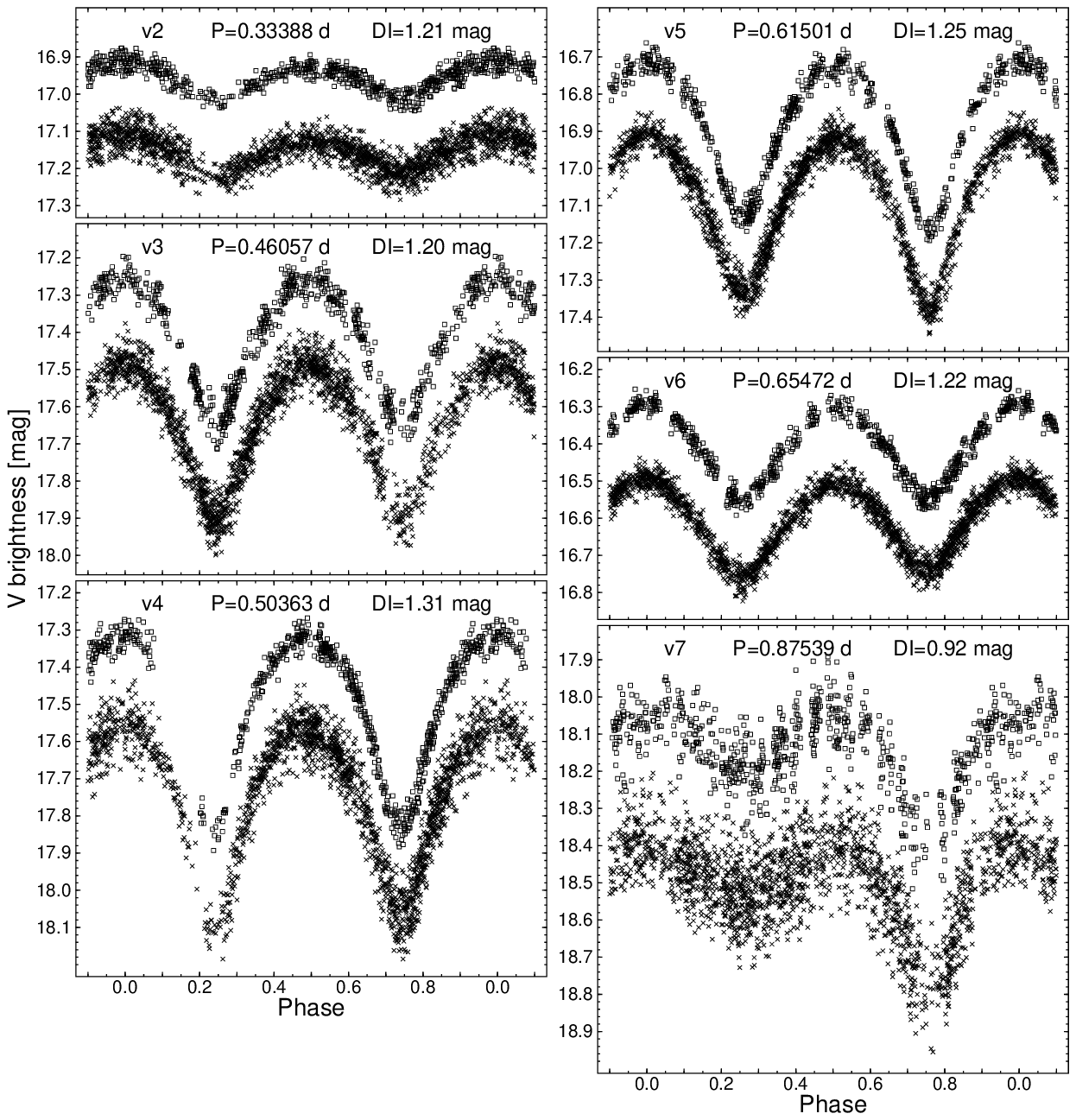}%
 Fig.\ \figwumalc. $V$ (crosses) and $I_{\rm C}$ (squares) light-curves of
 W UMa (v2 -- v6) and $\beta$ Lyr (v7) stars detected in the field of NGC\,7044.
 DI is a magnitude shift applied to the $I_{\rm C}$ data.
 The ordinate scale is the same in {\em all panels}.
\endfig

\vskip0.6\baselineskip plus1pt minus1pt
{\em \secVarWUMa. W UMa and $\beta$ Lyr binaries}
\vskip0.4\baselineskip plus1pt minus1pt

The $VI_{\rm C}$ light curves of five W UMa type (v2 -- v6) and one
$\beta$ Lyr type (v7) stars are presented in Fig.\ \figwumalc. As can
be seen from the figure, the dephts of the eclipses in v2 are significantly lower
than in other W UMa systems we observed.
%This is probably an effect of 
%the eclipses being partial due to the small orbital inclination.
The large scatter in the phase diagram of v7 comes from the fact that 
it is a very faint star ($V=18.5$ mag). The location of this variable
on the blue side of the main sequence in the color-magnitude diagram 
(see Fig.\ \figcmd) suggests that v7 is not a member of NGC\,7044. 

W UMa type binary systems are known to obey a relation between their 
period, color and absolute brightness (Ruci\'nski 1994, 2004): 
$$ M_V = -4.43\,\log(P/{\rm d}) + 3.63\,(V-I_{\rm C})_0 - 0.31,$$
where $(V-I_{\rm{C}})_0$ is the intrinsic color index measured at 
the maximum brightness of the binary system and $P$ is the orbital period.
The uncertainity of this calibration is approximately $\pm$0.35 mag in $M_V$
(Ruci\'nski and Duerbeck 1997). 
Because of this, the period-luminosity relation can be used 
only for deriving the distance to a large group of contact binaries
in one stellar cluster or galaxy (for a discussion see Ruci\'nski 2004). 
It can be also used as a tool 
for sieving out field interlopers from a sample of W UMa binaries in
a given open or globular cluster (see, for example, Ruci\'nski 2000).

\begintfig{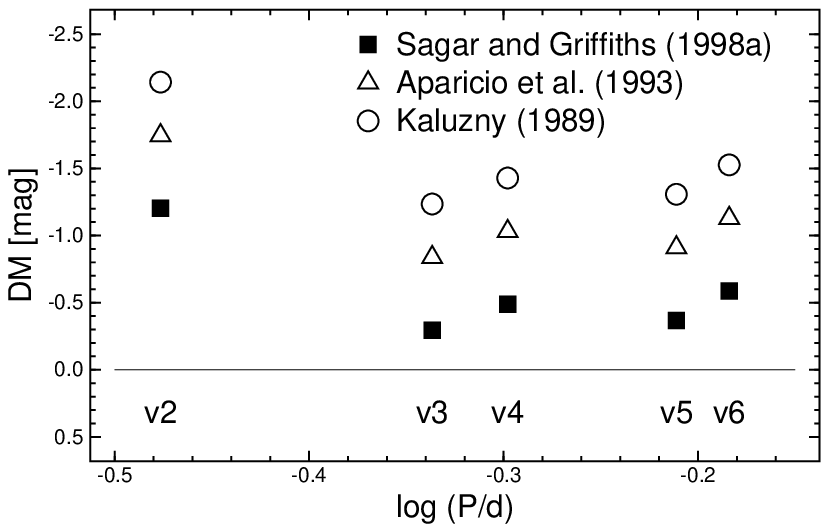}%
 Fig.\ \figwumaplr. Period-luminosity relation for W UMa type
 stars in the field of NGC\,7044. DM is the difference between absolute
 brightness calculated from the apparent $V$-magnitude and cluster
 parameters derived by Sagar and Griffiths (1998a, filled squares),
 Aparicio {\em et al}.\ (1993, triangles), and Ka\l{}u\.zny (1989, circles) and
 the magnitude calculated from Ruci\'nski's (1994) relationship.
\endfig

We compared pairs of absolute magnitudes of 
W UMa stars we observed: $M_{\rm{cal}}$ computed from the relation
given above and $M_{\rm{obs}}$ derived from the apparent 
$V$-magnitude and cluster distance modulus. We adopted 
distance moduli $(m-M)_V$ and reddenings $E(B-V)$ and
$E(V-I_{\rm C})$ from Sagar and Griffiths
(1998a), Aparicio {\em et al}.\ (1993), and Ka\l{}u\.zny (1989).
In calculations we assumed the reddening ratio given by Moro and Munari (2000),
$E(V-I_{\rm C})/E(B-V)=1.1$ mag. 
The differences $\Delta M = M_{\rm obs}-M_{\rm cal}$ are plotted
against the logarithm of orbital period in Fig.\ \figwumaplr.

From Fig.\ \figwumaplr\ one can see a large spread of distance moduli (and reddenings)
determined for the cluster by different authors. Only photometry of
Sagar and Griffiths (1998a) agrees reasonably well (within limits of 
accuracy of the Ruci\'nski's (1994) calibration) with the
period-luminosity relation.
More importantly, among each group of brightness differences
(defined by a set of cluster parameters),
stars v3, v4, v5, and v6 have similiar values of $\Delta M$, while the value for v2
is approximately 0.9 mag smaller. Consistent results for v3 -- v6 allow us
to conclude that these binaries are members of NGC\,7044, whereas v2 is a 
foreground star. It should be also noted that all observed W UMa systems
are located on the main sequence in the color-magnitude diagram 
shown in Fig.\ \figcmd.

The mean distance modulus $(m-M)_V$ of NGC\,7044 derived from the four 
W UMa stars assumed to be cluster members is $14.2\pm0.1\pm0.35$ mag 
(where 0.1 mag is the r.m.s.\ error of the mean and 0.35 mag
is the uncertainty of luminosity calibration). Taking $E(B-V)=0.7$
mag and assuming $A_V/E(B-V)=3.0$ mag, this translates into
a distance of 2.4 kpc, smaller than 
previous estimates but still within the range of uncertainty.

\begintfig{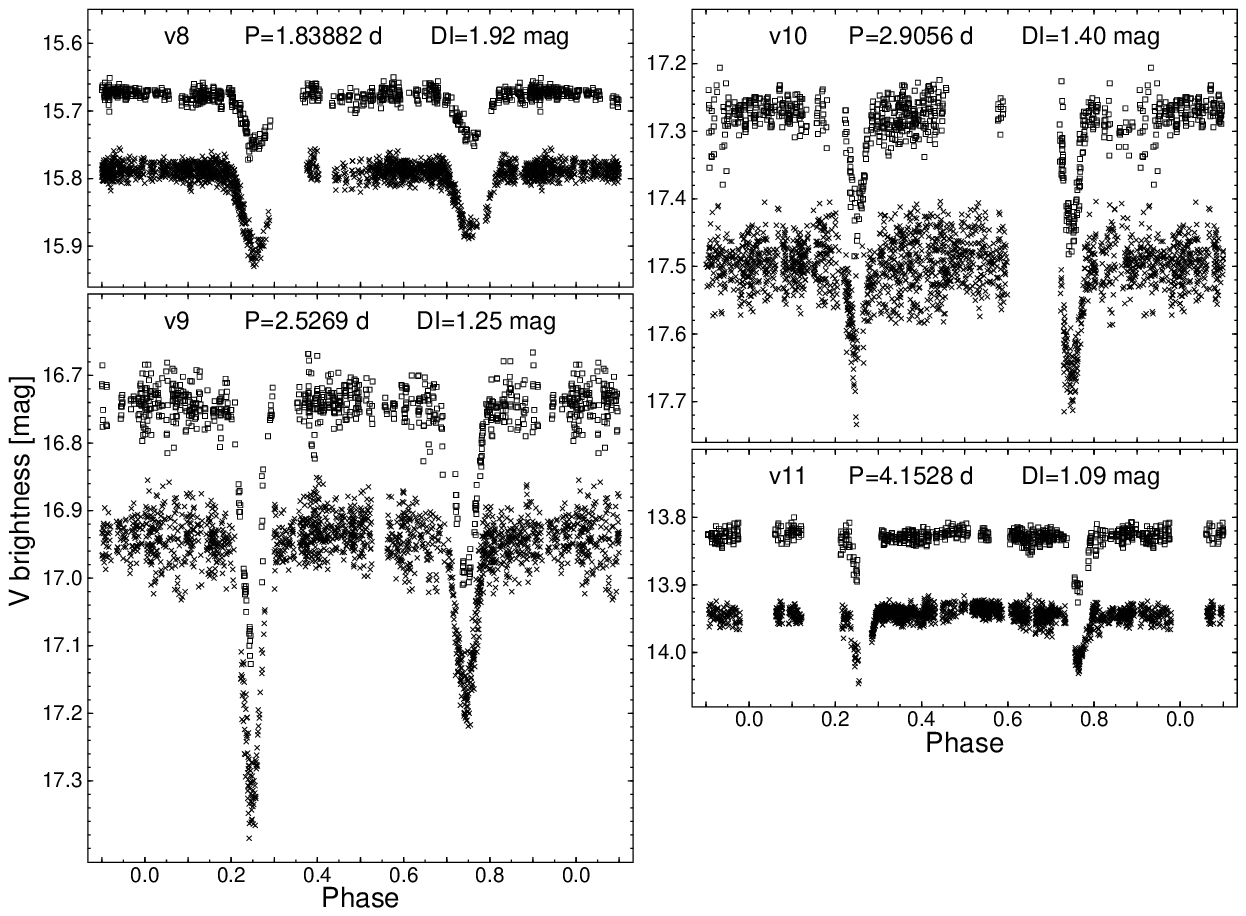}%
 Fig.\ \figdeblc. $V$ (crosses) and $I_{\rm C}$ (squares) light-curves of
 $\beta$ Per stars in the field of NGC\,7044 for which orbital period could be determined.
 DI is a magnitude shift applied to the $I_{\rm C}$ data.
 The ordinate scale is the same in {\em all panels}.
\endfig

\vskip0.6\baselineskip plus1pt minus1pt
{\em \secVarBPer. $\beta$ Per binaries}
\vskip0.4\baselineskip plus1pt minus1pt

Six other eclipsing systems discovered in the field of NGC\,7044, namely 
v8 through v13, we classify as $\beta$ Per stars. These binaries
show light variations typical for detached eclipsing 
binaries, that is, relatively sharp eclipses with constant brightness
between them. We were able to determine orbital periods for only four of
these stars (v8 -- v11). Their light curves are presented in Fig.\
\figdeblc.

\begintfig{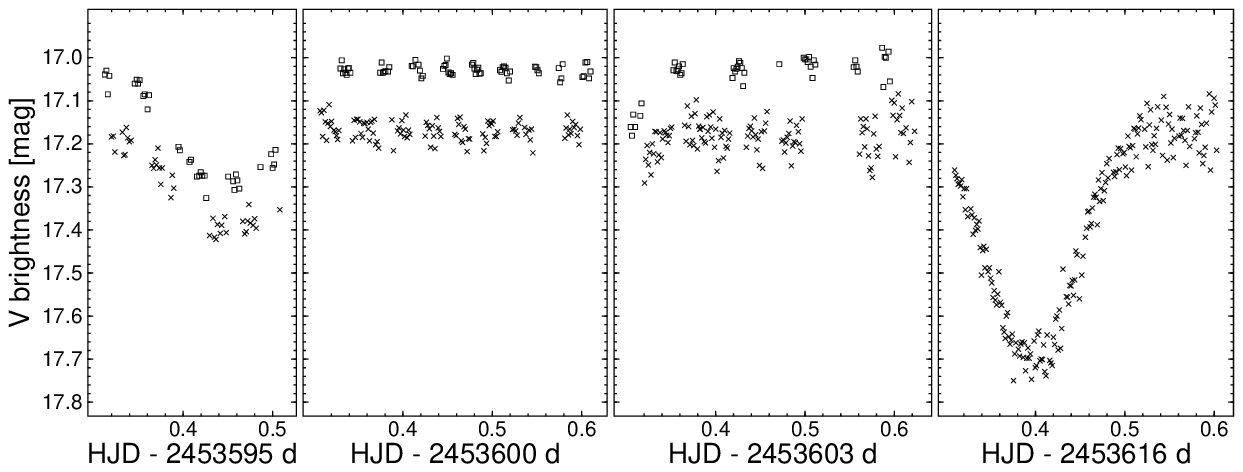}%
 Fig.\ \figdebvxii. $V$ (crosses) and $I_{\rm C}$ (squares) observations
 of v12 on four nights showing evidence for three eclipses ({\em first}, 
 {\em third\/} and
 {\em fourth panel\/}). The {\em second panel} shows an example of a night outside 
 eclipse. $I_{\rm C}$ magnitudes were shifted by 1.3 mag.
\endfig

\begintfig{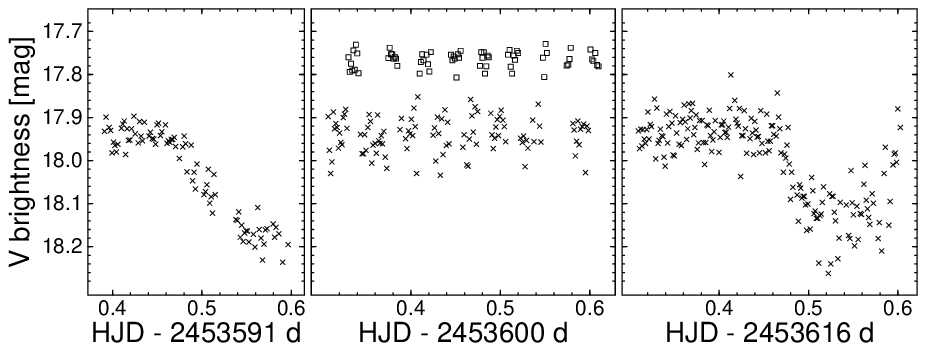}%
 Fig.\ \figdebvxiii. $V$ (crosses) and $I_{\rm C}$ (squares) observations
 of v13 on three nights showing evidence for two eclipses ({\em first\/} and {\em third
 panel\/}). The {\em second panel} shows an example of a night outside eclipse.
 $I_{\rm C}$ magnitudes were shifted by 1.5 mag.
\endfig

\begintfig{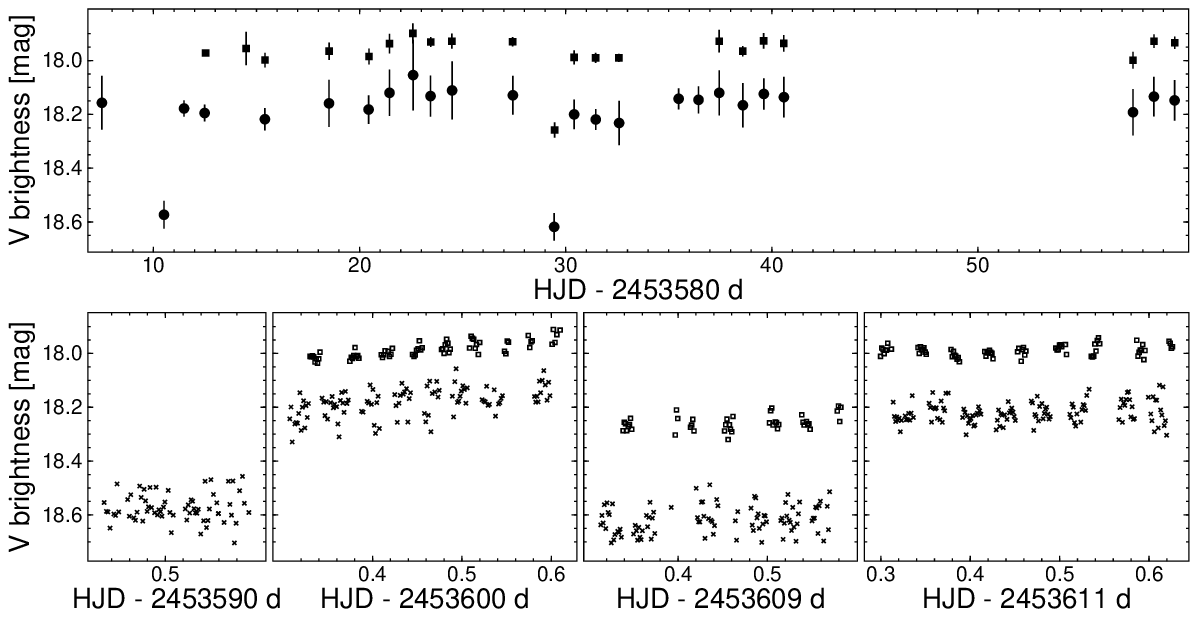}%
 Fig.\ \figdebvxiv. $V$ (crosses) and $I_{\rm C}$ (squares) observations
 of v14. The {\em upper panel} shows average nightly magnitudes with error bars 
 of length equal to the rms errors. The {\em bottom panels} present
 four nights of observation with indication for two eclipses ({\em first\/} and {\em third
 panel\/}). $I_{\rm C}$ magnitudes were shifted by 1.75 mag.
\endfig

Four nights of observations of the eclipsing variable v12 are shown in Fig.\
\figdebvxii. For this star, we observed only two eclipses with clearly 
defined minimum of brightness ({\em first} and {\em fourth panels\/} of Fig.\ \figdebvxii)
and indication for another eclipse on HJD 2453603 ({\em third panel\/} of 
Fig.\ \figdebvxii).
The time span $dt$ between secondary ({\em first panel\/} of Fig.\ \figdebvxii) and 
primary eclipses ({\em fourth panel\/} of Fig.\ \figdebvxii) is about 20.95 d.
Assuming circular orbit, the period would be $P=dt/(n+0.5)$ with $n$
being an integer number. Only $n=0$ folds the data into a satisfactory phase diagram.
Thus, the period of v12 would be very long, amounting to about 41.9 d. However,
if the eclipse, of which a part is shown in the {\em third panel\/} of Fig.\ \figdebvxii,
were a primary eclipse, we could estimate the orbital period from the two primary
minima with the time span of 13.17 d. In this case, the only possible period 
would be 13.17 d. The results of the two approaches exclude each other. This 
indicates that the eccentricity of the system is not zero. 
Further observations 
of this star are needed for a successful determination of its orbital period.

For the eclipsing star v13 we observed only two minima with 
the time difference of aproximately 25 d. They are shown in the {\em first\/} and 
{\em third panels\/} of Fig.\ \figdebvxiii. It is not certain of which type are 
these eclipses. Most probably, one is the primary and the other
is the secondary. Assuming that they are separated in phase by 0.5, only
long periods (50.0 and 16.67 d) give satisfactory phase diagram with no night
of constant brightness falling into any of the eclipses. It is
possible, however, that this is an eccentric binary system.

Another variable star which we tentatively classify as a $\beta$ Lyr system is v14. 
The epochs of the start and the end of eclipses were not observed for
this star, but the data show two eclipses, each lasting at least 5 hours.
The mean nightly magnitudes of this star are shown in the {\em upper panel\/} of Fig.\
\figdebvxiv, whereas {\em bottom panels\/} of the same figure show four nights with
indication for two eclipses ({\em first\/} and {\em third panels\/}).
There are variations of brightness outside eclipses (see {\em second bottom
panel\/} of Fig.\ \figdebvxiv) indicating the presence of proximity 
effects in the system. We could not determine a period for v14.

\begintfig{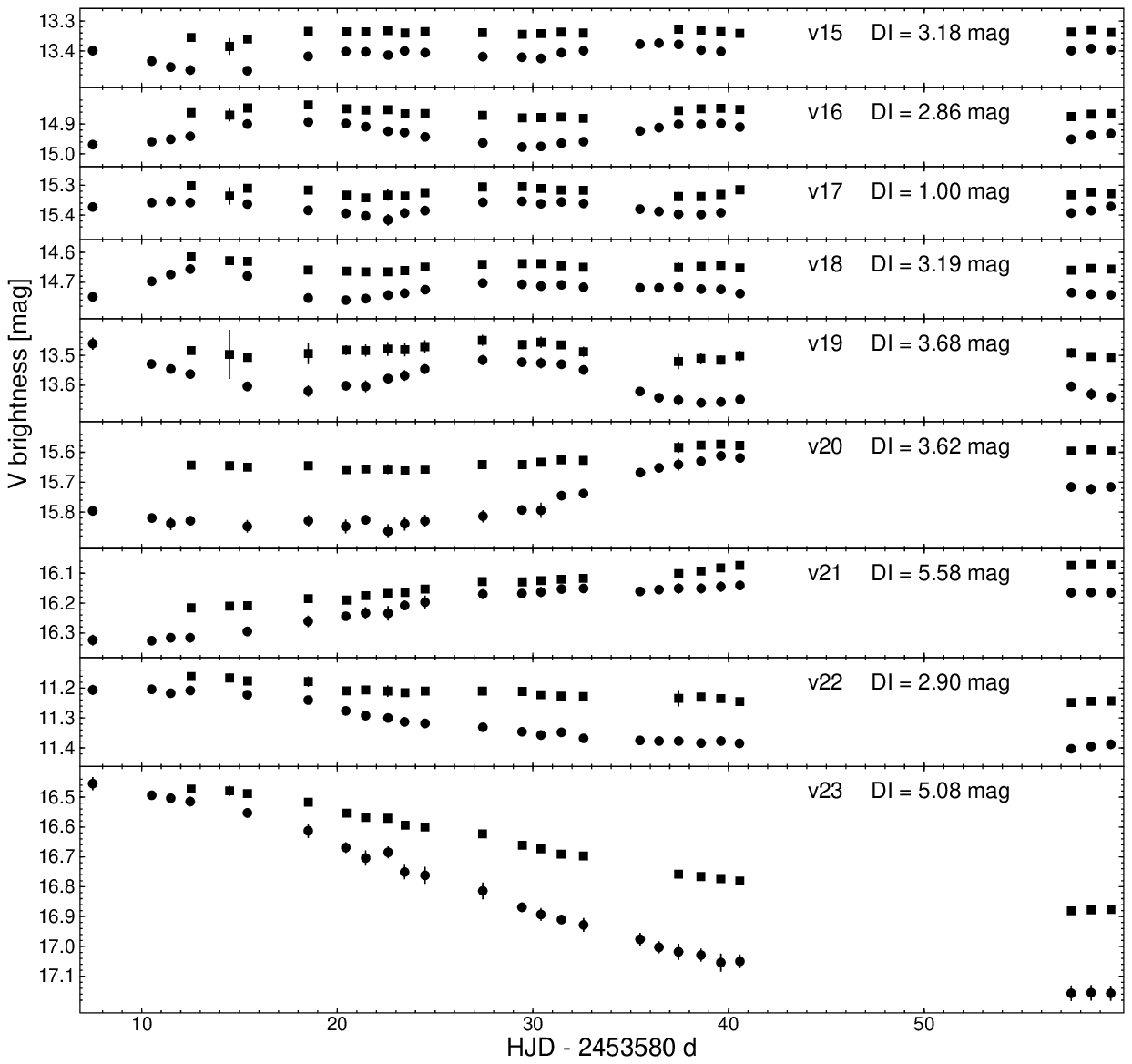}%
 Fig.\ \figrvlc. $V$- (filled crosses) and $I_{\rm C}$-filter (filled squares) mean nightly
 magnitudes of the red variable stars in the field of NGC\,7044.
 The error bars have a length of the rms errors.
 DI is a magnitude shift applied to the $I_{\rm C}$ data.
 The ordinate scale is the same in {\em all panels}.
\endfig

\vskip0.6\baselineskip plus1pt minus1pt
{\em \secVarIrr. Irregular stars}
\vskip0.4\baselineskip plus1pt minus1pt

The largest group of variable stars discovered in the field of NGC\,7044
consists of nine irregular variables (v15 -- v23). Their light curves 
are plotted in Fig.\ \figrvlc. Eight of these variables are red, 
and one (v17) is situated on the main sequence, above the cluster's turn-off 
point (see Fig.\ \figcmd). 

\vskip0.7cm
\centerline{\bf \secCMD. Color-Magnitude Diagram}
\vskip0.5cm

The $V$ vs.\ $(V-I_{\rm C})$ color-magnitude diagram for the
observed field of NGC\,7044 using all stars with realiable photometry
is shown in Fig.\ \figcmd{}a. The two clearly visible features 
can be seen in this diagram: a relatively wide
main sequence and a clump of red giants slightly above and about 0.5 mag 
to the red off the turn-off point. There is also a group of stars brighter
and bluer than turn-off point, indicating that some of them may be blue
stragglers.

\begintfig{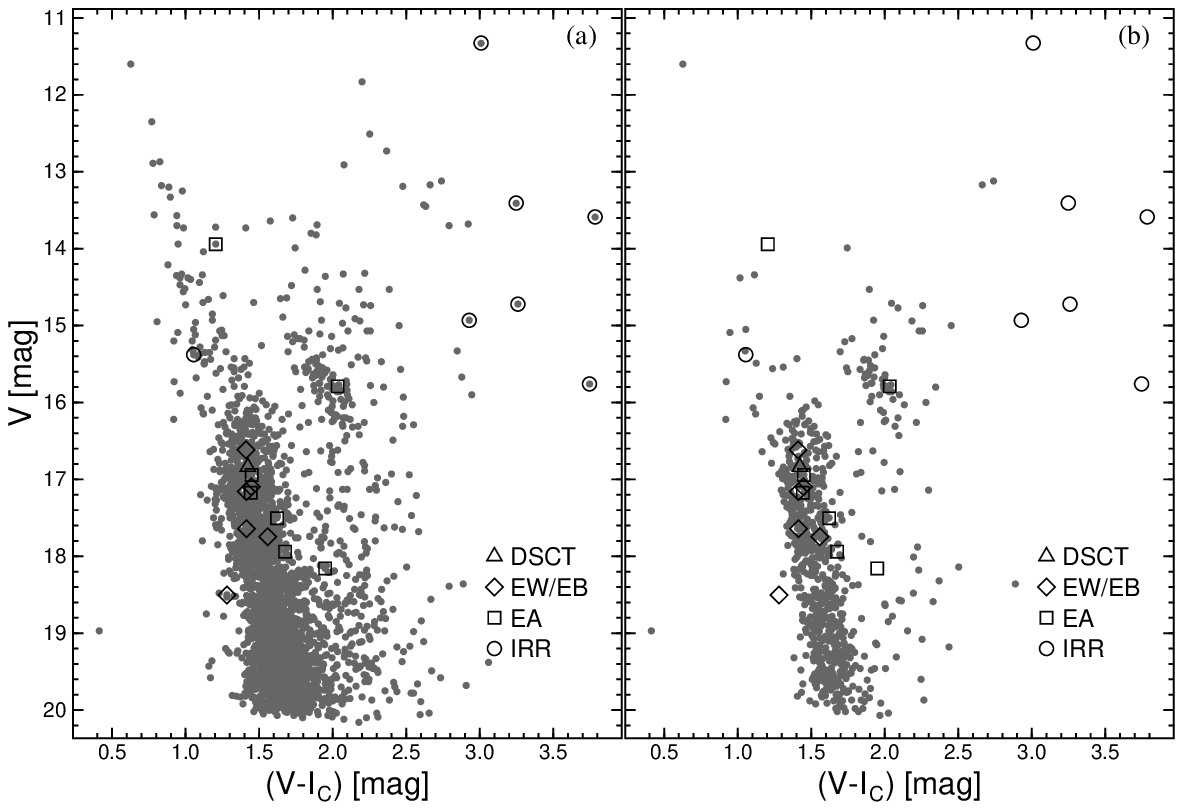}%
 Fig.\ \figcmd. The $V$ vs.\ $(V-I_{\rm C})$ color-magnitude
 diagrams for NGC\,7044: (a) using all observed stars
 with realiable photometric data, and (b) using only stars
 with the distance from cluster center smaller 
 than the derived cluster radius of 2.5 arcmin.
 Variable stars are indicated with open symbols,
 one $\delta$ Scuti star, with a triangle, W UMa and $\beta$ Lyr 
 binary systems (EW/EB), with diamonds, $\beta$ Per systems (EA), with
 squares, and irregular variable stars with circles 
 (the two extremely red stars, v21 and v23, are not shown).
 The ordinate and abscissa scales of the {\em left panel\/} are the same
 as the corresponding scales of the {\em right panel\/}.
\endfig

Since NGC\,7044 is very close to the Galactic plane ($b\approx4^\circ$),
the contamination by cluster non-members is
rather high, especially in regions well away from the cluster center. 
This manifests itself in Fig.\ \figcmd{}a mostly through the wide main sequence 
and a numerous group of faint stars on its red side.
In order to make a rough cleaning of color-magnitude diagram 
from possible non-members, we determined the cluster angular
diameter from the radial profile of star counts. We applied the same
procedure as that described by Sagar and Griffiths (1998b) in which
the cluster's center and diameter are derived iteratively. This analysis
resulted in a radius of about 2.5 arcmin.

The cleaned color-magnitude diagram was constructed using only 
stars with distance from the cluster center smaller than the estimated cluster radius.
It is given in Fig.\ \figcmd{}b. As expected, the cluster main sequence
and red giant clump are better defined in this figure in comparison 
to the original diagram in Fig.\ \figcmd{}a.

We also show in Fig.\ \figcmd\ the location of variable stars we detected. 
An information on their possible membership can be inferred from 
their position in this diagram. The membership of W UMa and $\beta$ 
Lyr stars was already discussed in Chapter 2.2. Among $\beta$ Per systems,
two stars (v9 and v12) are located on the cluster main-sequence,
two other stars (v10 and v13) are at the red border of the
main sequence, and v8 is in the red giant region. These stars are
very probable cluster members. Only v11 in this
group seems to be a non-member. A possible $\beta$ Lyr system v14 is a
foreground object.

Recently, Dias {\em et al}.\ (2006) determined membership probabilities
for 150 stars in the field of NGC\,7044 using the UCAC2 catalogue of
accurate positions and proper motions published by Zacharias {\em et al}.\
(2004). Unfortunately, the limiting $V$-brightness of this catalogue is 16
mag. As can be seen from Fig.\ \figcmd, this is well above the cluster
turn-off point. Because of that, most stars in the sample
studied by Dias {\em et al}.\ (2006) are
cluster non-members. In fact, the average proper motions of the assumed
members (35 stars) and non-members (115) derived by Dias {\em et al}.\ (2006) are
almost the same, so that distinguishing the two populations is not possible.
Thus, in our opinion, membership probabilities derived by these authors
are rather doubtful. 

Among stars analyzed by Dias {\em et al}.\ (2006) we find five variable stars,
v8 (membership probability 12 \%), v11 (41 \%), v16 (61 \%), v17 (3 \%), and
v21 (86 \%). These values mostly do not agree with
information on membership derived from our photometry. The very evident example
is the red variable v21, which is a highly probable cluster member according to
Dias {\em et al}.\ (2006), but from the photometric point of view it is definitely
not a member.

\vskip0.7cm
\centerline{\bf \secRMFit. Reddening Map and Cluster Parameters}
\vskip0.5cm

In order to verify variable reddening in NGC~7044, we calculated
a~map of reddening in central region of the observed field
using the procedure described by Pigulski and Ko{\l}aczkowski (1998).
The method assumes that all measured extinction is caused by
matter lying between the cluster and the observer and that it changes
smoothly over the field. These assumptions imply that differences in
color of member stars lying very close to each other are
intrinsic, and not due to differences in reddening.

We selected a sample of very probable cluster members, that is, stars
from or close to the cluster main sequence, with distance from the cluster
center $r<{}$2.5 arcmin, brightness $V>{}$16.7 mag, and $(V-I_{\rm C})$
colors from 1.3 to 1.8 mag. For a fiducial line used in the color excess
calculations, we adopted the isochrone of Girardi {\em et al}.\ (2000) with
$\log(\tau/\rm{yr})={}$9.1, shifted in color by 1 mag and in brightness
by 15.1 mag. The assumed scale of color changes was 25~pixels,
corresponding to about 0.2~pc at the cluster distance. In an iterative process we first calculated
the reddening map (an average color excess as a function of position
in the observed field) and next rejected stars with color excesses deviating
from the average value, counting these stars as probable cluster non-members.

The reddening map is shown in the {\em right panel\/} of
Fig.\ \figcmdredmap. The $E(V-I_{\rm C})$ ranges from
0.83 to 1.13 mag; the average value is 0.90 mag.
The color-magnitude diagram corrected for the effect of variable reddening is
presented in the {\em left panel\/} of Fig.\ \figcmdredmap. In comparison with Fig.\ \figcmd,
the main sequence is thinner and the turnoff point better defined.

\begintfig{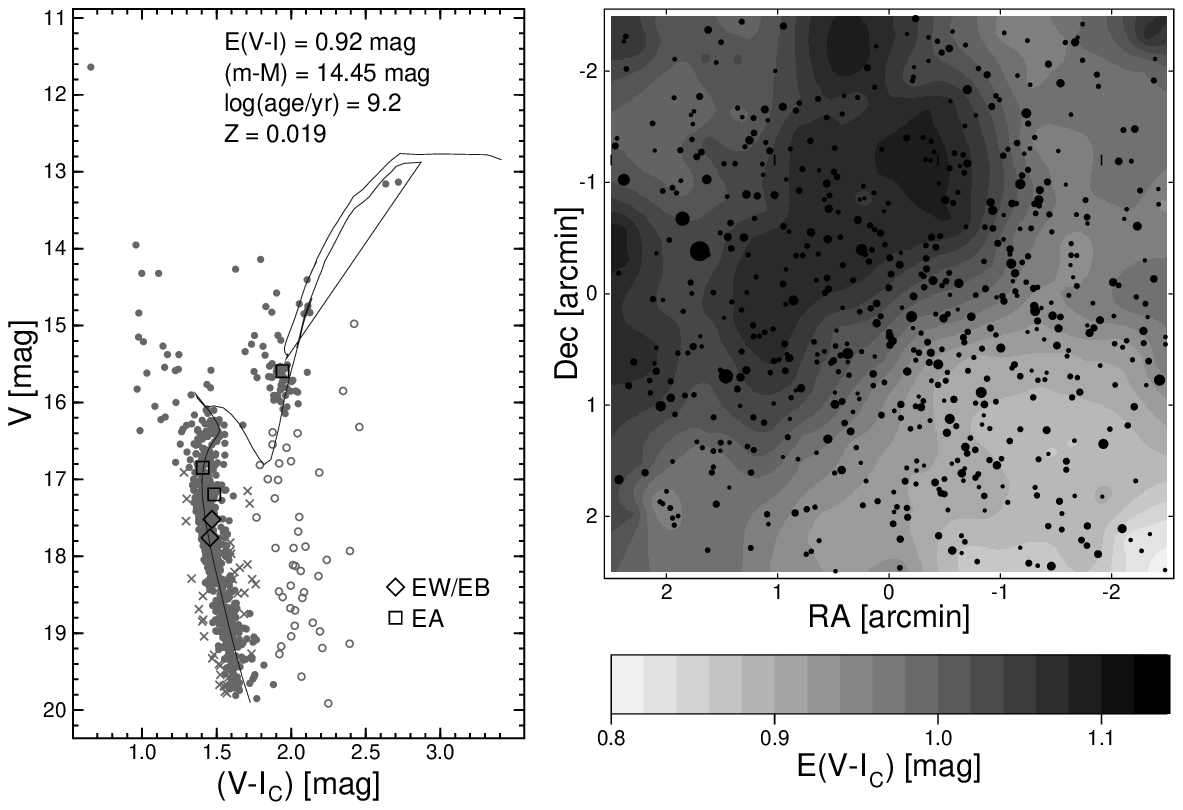}%
 Fig.\ \figcmdredmap. The $V$ vs.\ $(V-I_{\rm C})$ color-magnitude
 diagram of NGC\,7044 corrected for the effect of variable
 reddening ({\em left panel}) and the reddening map ({\em right panel})
 showing the distribution of the $E(V-I_{\rm C})$ color excess across the cluster field.
 The reddening map covers the central 5${}\times{}$5 arcmin$^{\rm 2}$
 of NGC\,7044. Only stars from this area are shown in the
 color-magnitude diagram. Very probable field stars are indicated with open
 circles and stars found in the reddening map derivation to be cluster non-members
 are shown with crosses.
 The best-fitting isochrone of Girardi {\em et al}.\ (2000) for
 the marked values of color excess, distance modulus and age is shown as
 solid line.
 Several variable stars are indicated with open symbols,
 W UMa and $\beta$ Lyr binary systems (EW/EB) with diamonds,
 $\beta$ Per systems (EA) with squares.
 The ordinate and abscissa scales of the {\em left panel\/} are the same
 as the corresponding scales in Fig.\ \figcmd.
\endfig

We attempted fitting of the theoretical isochrones by Girardi {\em et al}.\ (2000)
to the corrected color-magnitude diagram by eye. These isochrones include the
effects of core overshooting and mass loss on the red giant branch. We assumed
solar metallicity. The best fit was achieved for the following set of parameters:
$E(V-I_{\rm C})={}$0.92 mag, $(m-M)_V={}$14.45 mag and
$\log(\tau/\rm{yr})={}$9.2. The isochrone corresponding to the above parameters
is shown with the solid line in the {\em left panel\/} of Fig.\ \figcmdredmap.
As can be seen from the figure, the isochrone does not fit the group of red giants very well.
The same problem was encountered by previous investigators of the cluster.
It may be probably explained by the deficiencies of input
physics of theoretical stellar models.
Following the suggestion of Sagar and Griffiths (1998a), we also tried to
fit metal-poor isochrones to the color-magnitude diagram of NGC\,7044,
but the quality of fit was still unsatisfactory.  

Our cluster parameters are not very different from the values
obtained by Sagar and Griffiths (1998a). The distance modulus derived by us
is 0.15 mag smaller than the value of Sagar and Griffiths (1998a).
This was, however, to be expected because our
$VI_{\rm C}$ photometry is tied to the photometric data of Sagar and Griffiths (1998a)
through the transformation equations (see Chapter \secObsRed).
The small differences come probably from the fact that our color-magnitude diagram
has been corrected for the effect of variable reddening and from using a different
set of isochrones.
In comparison to distance modulus of Sagar and Griffiths (1998a)
our value from isochrone fitting is closer to that derived by us from
W UMa stars (Paragraph \secVarWUMa), but agreement is still far from
satisfactory.

It should be noted that Sagar and Griffiths (1998a) fitted isochrones
to the blue edge of the original broad main sequence and concluded
that a significant fraction of cluster members are binaries. Our
results indicate that the cluster main sequence corrected for differential
reddening can be reasonably well described with a population of
mostly single stars.

\vskip0.5cm
{\bf Acknowledgements.} 
%I want to express my gratitude to A.\ Pigulski
%for the careful reading of the manuscript and valuable comments.
This work was supported by Polish Ministry of Science 
grant N203\,014\,31/2650. We thank A.\ Pigulski for taking some
observations of NGC$\,$7044 used in this paper.

\vskip0.5cm
\centerline{REFERENCES}
\vskip0.3cm

\def\ref#1#2#3#4#5{#1\ #2, {\em #3}, {\bf #4}, #5.}

\everypar={\hangafter=1\hangindent=1cm}

\small

\parindent=0pt

\ref
{Aparicio, A., Alfaro, E.J., Delgado, A.J., Rodr\'\i{}guez-Ulloa, J.A.,
 and Cabrera-Ca\~no, J.}{1993}{Astron.\ J.}{106}{1547}

\ref
{Dias, W.S., Florio, V., Assafin, M., Alessi, B.S., and Libero V.}
 {2006}{Astron.\ Astrophys.}{446}{949}

\ref
{Girardi, L., Bressan, A., Bertelli, G., Chiosi, C.}{2000}{Astron.\ Astrophys.\ Suppl.\ Series}{141}{371}
	
\ref
{Jerzykiewicz, M., Pigulski, A., Kopacki, G.,
 Mia\l{}kowska, A., and Niczyporuk, S.}{1996}{Acta Astron.}{46}{253}

\ref
{Ka\l{}u\.zny, J.}{1989}{Acta Astron.}{39}{13} 

\ref
{Moro, D., and Munari, U.}{2000}{Astron.\ Astrophys.\ Suppl.\ Series}{147}{361}

\ref
{Pigulski, A., and Ko\l{}aczkowski, Z.}{1998}{Mon.\ Not.\ R.\ Astron.\ Soc.}{298}{753}

\ref
{Ruci\'nski, S.M.}{1994}{Pub.\ Astron.\ Soc.\ Pacific}{106}{462}

\ref
{Ruci\'nski, S.M.}{2000}{Astron.\ J.}{120}{319}

\ref
{Ruci\'nski, S.M.}{2004}{New Astron.\ Rev.}{48}{703}

\ref
{Ruci\'nski, S.M., and Duerbeck, H.W.}{1997}{Pub.\ Astron.\ Soc.\ Pacific}{109}{1340}

\ref
{Sagar, R., and Griffiths, W.K.}{1998a}{Mon.\ Not.\ R.\ Astron.\ Soc.}{299}{1}

\ref
{Sagar, R., and Griffiths, W.K.}{1998b}{Mon.\ Not.\ R.\ Astron.\ Soc.}{299}{777}

\ref
{Stetson, P.B.}{1987,}{Pub.\ Astron.\ Soc.\ Pacific}{99}{191}

\ref
{Zacharias, N., Urban, S.E., Zacharias, M.I., Wycoff, G.L.,
 Hall, D.M., Mo\-net, D.G., and Rafferty, T.J.}{2004}{Astron.\ J.}{127}{3043}

\vfill\eject\end